\def\luv{\Lambda_{UV}}

\def\li2{{\rm Li}_2}

\def\roughly#1{\,\,\raise.3ex\hbox{$#1$\kern-.75em\lower1ex\hbox{$\sim$}}\,\,}

\def\beq{\begin{equation}}
\def\eeq{\end{equation}}
\def\bea{\begin{eqnarray}}
\def\eea{\end{eqnarray}}
\def\bq{\begin{quote}}
\def\eq{\end{quote}}
\def    \missEt      {\ifmmode{/\mkern-11mu E_T}\else{${/\mkern-11mu E_T}$}\fi}
\def\BReq {\ifmmode { {\cal B}_{eq}} \else {${\cal B}_{eq}$}\fi}
\def\BRR  {\ifmmode { {\cal B}_{R}} \else {${\cal B}_{R}$}\fi}

\def\gappeq{\mathrel{\rlap {\raise.5ex\hbox{$>$}}
{\lower.5ex\hbox{$\sim$}}}}

\def\lappeq{\mathrel{\rlap{\raise.5ex\hbox{$<$}}
{\lower.5ex\hbox{$\sim$}}}}

\def\bbz{fa Z \kern-8.9pt Z}
\documentstyle [12pt,epsfig]{article}
\begin{document}
\thispagestyle{empty}
\vspace*{-1cm}
\begin{flushright}
{CERN-TH/97-145} \\
{hep-ph/9706540} \\
\end{flushright}
\vspace{1cm}
\begin{center}
{\large {\bf Extracting Supersymmetry-Breaking Effects from \\ Wave-Function 
Renormalization}} \\
\vspace{.2cm}                                                  
\end{center}

\begin{center}
\vspace{.3cm}
{\bf 
G.F. Giudice\footnote{On leave of absence from INFN, Sez. di Padova, Italy.} }
 and                                                     
{\bf R. Rattazzi}
\\                          
\vspace{.5cm}
{ Theory Division, CERN, CH-1211, Gen\`eve 23, Switzerland} \\
\end{center}                              

\vspace{2cm}

\begin{abstract}
We show that in theories in which supersymmetry breaking is communicated by 
renormalizable perturbative interactions, it is possible to extract
the soft terms for the observable fields from wave-function renormalization.
Therefore all the information about soft terms
can be obtained from anomalous dimensions and
$\beta$ functions, with no need to further compute any Feynman diagram.
This method greatly simplifies calculations which are rather
involved if performed in terms of component fields. For illustrative
purposes we reproduce known results of theories with gauge-mediated
supersymmetry breaking. We then use our method to obtain new results of
phenomenological importance. We calculate the next-to-leading correction to
the Higgs mass parameters, the two-loop soft terms induced by messenger-matter
superpotential couplings, and the soft terms generated by messengers
belonging to vector supermultiplets.
\end{abstract}
\vfill 
\begin{flushleft}
CERN-TH/97-145\\
June 1997\\
\end{flushleft}

\newpage

\section{Introduction}

Theories in which supersymmetry breaking is mediated
by renormalizable perturbative interactions have an interesting advantage
over the gravity-mediated scenario. This is because in these
theories the soft terms are in priciple calculable quantities,
very much like $g-2$ in QED. Gauge-mediated models~\cite{gauge,din}, 
in particular,
have also a strong motivation as they elegantly
solve the supersymmetric flavour problem. In the simplest version of these
theories, the gaugino and sfermion masses arise respectively from one- and
two-loop 
finite diagrams. Their evaluation, while conceptually straightforward,
takes in practice some effort. With the increasing number of non-minimal 
examples it would certainly be very useful to have a quick and reliable
method to evaluate soft masses. When scanning through models it is
important to know right away if the predicted masses are
phenomenologically
consistent and whether there are any interesting predictions. It is the
purpose of this paper to develop one such technique and to apply it to
a number of interesting cases. These include theories in which gauge as
well as Yukawa interactions mediate soft masses and also theories in
which the messengers are massive gauge supermultiplets.

We will be interested in the case in which the breaking of supersymmetry
and the scale of its mediation to the observable sector
are completely determined
by the scalar and auxiliary component  vacuum expectation values (VEV)
of a chiral superfield $\langle X\rangle =M+\theta^2 F$. 
In the case of conventional
gauge mediation $X$ gives mass to the messengers via the superpotential
coupling $X\Phi {\bar \Phi}$. Moreover we are interested in the regime
$F\ll M^2$
for which, in the effective field theory below the messenger scale
$M$, supersymmetry
is broken {\it only} by soft effects. This allows us to use a manifestly
supersymmetric formalism to keep track of supersymmetry-breaking 
effects as well. 
Due to the non-renormalization of the superpotential,
the  gaugino and sfermion masses  arise just from $X$-dependent
renormalizations of respectively the gauge and matter kinetic terms,
{\it i.e.} from the gauge charge $g^2$ and matter wave-function
renormalizations
$Z_i$. In the presence of only one spurion
$X$, the kinetic functions can be obtained by
calculating first $g^2$ and $Z_i$ in the supersymmetric limit
in which $X=M$ is just a c-number mass, 
and by substituting later $M$ with the superfield $ X$. This is the
crucial point of our paper. The $g^2(M,\mu)$ and $Z_i(M,\mu)$ renormalized
at a
 scale $\mu\ll M$ are simply evaluated by solving the Renormalization
Group (RG) equations. At the end, the substitution $M\to X$ in $g^2$ is made
since
this
quantity, at least at one-loop, has to be holomorphic. On the other hand,
the only substitution in $Z_i$ consistent with the chiral
reparametrization
$X\to e^{i\phi}X$ is given by $M\to \sqrt{XX^\dagger}$. As it will be
shown
below, a very interesting advantage of this method is that the two-loop 
sfermion masses of gauge mediation are determined only by the
one-loop RG equations. 

This paper is organized as follows. Section ~2 contains the essence of
our paper. There we  describe our method, its regime of validity,
and give the general formulae relating the supersymmetry-breaking terms 
to $X$-derivatives of wave-function renormalizations.  For illustration,
we
reproduce the known soft terms of gauge-mediated theories. 
Later sections are devoted to some
applications of phenomenological interest.
In sect.~3 we compute the next-to-leading corrections to the Higgs mass
parameters, while soft terms induced by superpotential couplings between
messengers and matter are studied in sect.~4. The case of messengers 
belonging to vector supermultiplet is discussed in sect.~5, and sect.~6
contains a summary of our results.
 
\section{The Method and its Application to Gauge Mediation}

In this section we will derive the general formulae which relate the
soft terms to $X$-derivatives of wave-function renormalizations. 
In our formulation $X$ is a background superfield which has
both a scalar and an auxiliary VEV, $\langle X\rangle
 =M+\theta^2 F$. Here $M$ determines the mass scale of the messenger
fields which communicate supersymmetry-breaking to the observable fields.
For instance, in gauge-mediated theories,
the messenger sector consists of $N$ pairs of chiral
superfields $\Phi$, $\bar \Phi$ getting mass from the superpotential
interaction to $X$
\beq
\int d^2\theta X \Phi \bar\Phi ~.
\label{massmess}
\eeq

The gaugino mass renormalized at a scale $\mu$ is 
determined by the $X$-dependendent wave function $S$ of the gauge multiplet,
\beq
{\cal L}=\int d^2\theta SW^{a\alpha}W^a_\alpha + {\rm h.c.}
\label{gwave}
\eeq
Here $a$ and $\alpha$ are gauge and spinorial indices, respectively. At
the one-loop order, to which we are now interested, $S$ is a holomorphic 
function $S(X,\mu)$.
Then at the minimum $\langle X\rangle
 =M+\theta^2 F$, we obtain that the supersymmetry-breaking
gaugino mass is given by
\beq
{\tilde M}_g (\mu )=\left. -\frac{1}{2}\frac{\partial \ln S(X,\mu)}
{\partial \ln X}\right|_{X=M}~\frac{F}{M}~.
\label{gaugino}
\eeq
The $S$ functional dependence on the Goldstino superfield $X$ can be 
obtained by observing that the scalar component of $S$ is
\beq
S(M,\mu)=\frac{\alpha(M,\mu)^{-1}}{16\pi}-\frac{i\Theta}{32 \pi^2}~,
\label{s}
\eeq
where $\alpha$ is the gauge-coupling strength and $\Theta$ the topological
vacuum angle. Now, the function $\alpha(M,\mu)$ is determined by integrating
the RG evolution from an ultraviolet scale $\Lambda_{UV}$ down to
$\mu$ across the messenger threshold $M$.
The one-loop RG equation is
\beq
\frac{d}{dt}\alpha^{-1}=\frac{b}{2\pi}~,
\label{runalp}
\eeq
where $t=\ln \mu$ and  $b$ is the $\beta$-function coefficient, 
given by $b=3N_c-N_f$
in an $SU(N_c)$ gauge theories with $N_f$ flavours.
Above the scale $M$, the $N$ messenger superfields transforming as $N_c+
\bar{N_c}$ under $SU(N_c)$ contribute to the gauge running and the
$\beta$-function coefficient is 
\beq
b^\prime= b-N~,
\eeq
if $b$ is the coefficient below $M$. By using tree level matching and one-loop
running one gets
\beq
{\rm Re}~S(M,\mu)=\frac{\alpha^{-1}(M,\mu)}{16\pi}=
\frac{\alpha^{-1}(\Lambda_{UV})}{16\pi}+
\frac{b^\prime}{32\pi^2}\ln \frac{|M|}{\Lambda_{UV}}+\frac{b}{32\pi^2}\ln
\frac{\mu}{|M|}~.
\label{aM}
\eeq
This expression and the requirement of holomorphy fix $S(X,\mu)$ to have the
form 
\beq
S(X,\mu)=S(\luv )+\frac{b^\prime}{32\pi^2}\ln
\frac{X}{\luv}+\frac{b}{32\pi^2}\ln
\frac{\mu}{X}~.
\label{sx}
\eeq
Notice that the above equation reproduces also the renormalization 
$\Theta\to \Theta +(b-b^\prime){\arg}(X)$ as dictated by the chiral
anomaly. 
By differentiating eq.~(\ref{sx}) with respect to $\ln X$ keeping
$S(\luv )$ fixed, we can express the gaugino mass in
eq.~(\ref{gaugino}) as
\beq
{\tilde M}_g (\mu)=\frac{\alpha(\mu)}{4\pi}~N~\frac{F}{M}.
\label{gaugau}
\eeq
This is the familiar expression obtained by explicit loop integration.
Notice that it already contains its one-loop RG evolution.

We now turn to discuss the supersymmetry-breaking terms for the
matter fields. 
Let $Z_Q$ be the wave-function renormalization of
the chiral superfield $Q$
\beq
{\cal L}=\int d^4\theta Z_Q(X,X^\dagger )~Q^\dagger Q~.
\label{lags}
\eeq
In contrast to the case of the gauge multiplet, here we are dealing with
the renormalization of a $D$-term and therefore $Z_Q$ is a real function
of {\it both} $X$ and $X^\dagger$. 
 Replacing the superfield-$X$ VEV in
eq.~(\ref{lags}), we find
\beq
{\cal L}=\int d^4\theta \left.
\left( Z_Q+\frac{\partial Z_Q}{\partial X}F\theta^2
+\frac{\partial Z_Q}{\partial X^\dagger}F^\dagger{\bar \theta}^2
+\frac{\partial^2 Z_Q}{\partial X\partial X^\dagger}FF^\dagger\theta^2{\bar
\theta}^2\right) \right|_{X=M}
~Q^\dagger Q~.
\label{lagss}
\eeq
It is useful to define a new variable $Q^\prime$ with canonically 
normalized kinetic terms,
\beq
Q^\prime \equiv Z_Q^ {\frac{1}{2}} \left. \left (1+
\frac{\partial \ln Z_Q}{
\partial X}F\theta^2\right) \right|_{X=M} ~Q~\equiv {\cal Z}Q.
\label{qp}
\eeq
Expressing eq.~(\ref{lagss}) in terms of $Q^\prime$, we find that linear
terms in $F$ disappear, and we are left with a
quadratic term corresponding to a supersymmetry-breaking mass for the
scalar component of the chiral multiplet
\beq
{\tilde m}_Q^2(\mu )=\left. -\frac{\partial^2 \ln Z_Q(X,X^\dagger , \mu)}
{\partial \ln X ~\partial \ln X^\dagger}\right|_{X=M}
~\frac{FF^\dagger}{MM^\dagger}~.
\label{squak}
\eeq
The derivatives in eq.~(\ref{squak}) are computed keeping the  
couplings at $\luv $ fixed.
Again we have made explicit the dependence on the renormalization scale $\mu$.
The redefinition of the chiral superfield $Q$ in eq.~(\ref{qp}) however
does not leave the superpotential invariant, and in particular it gives
rise to $A$-type supersymmetry-breaking terms proportional to $F$. Considering
a superpotential $W(Q)$ with the fields $Q_i$ redefined by $Q^\prime_i=
{\cal Z}_i Q_i$, we obtain an $A$-type
 contribution to the scalar potential
\beq
V=\sum_iA_iQ_i\partial_{Q_i}W(Q)+{\rm h.c.}
\label{pot}
\eeq
involving the scalar components of the corresponding superfields, where 
\beq
A_i(\mu )=\left. \frac{\partial \ln Z_{Q_i}(X,X^\dagger,\mu)}
{\partial \ln X}\right|_{X=M}~\frac{F}{M}~.
\label{aa}
\eeq
Notice that eq.~(\ref{aa})
corresponds to the coefficients $A_i$ defined in eq.~(\ref{pot}) when $W(Q)$ is 
expressed in terms of renormalized fields and running coupling constants
at the scale $\mu$.

Let us now obtain $Z_Q(X,X^\dagger)$ from RG evolution. The crucial remark
is that $Z_Q$ must be a function of the product $XX^\dagger$. This is 
due to invariance under the chiral symmetry $X\to e^{i\varphi}X$, 
$\bar \Phi\Phi\to e^{-i\varphi}\bar \Phi\Phi$. This symmetry is
anomalous, but this has no effect in perturbation theory as the anomaly
only affects the $\Theta$ angle, as seen in eq.~(\ref{sx}). It is then
straightforward to solve the RG evolution for the c-number $Z_Q(M,\mu)$
and substitute $M\to \sqrt{XX^\dagger}$ at the end of the calculation.
Now, $Z_Q$ will just be a power series in  $L_\Lambda=\ln (\mu^2/\luv^2)$ and 
$L_X=\ln(\mu^2/XX^\dagger)$. This property has a very important consequence
that greatly simplifies the calculation of soft masses. 
The contribution to
the wave-function renormalization 
at the loop order $\ell$
can be written as
\beq
\ln Z_Q(X,X^\dagger ,\mu )=\alpha_{UV}^{\ell -1} P_\ell (\alpha_{UV}L_X ,
\alpha_{UV}L_\Lambda ) ~,
\eeq
where $\alpha_{UV}=\alpha (\luv )$ and $P_\ell$ ia s function computed
by integrating the RG equations.
Using eq.~(\ref{squak}), we see that the corresponding contribution
to the soft scalar mass is
\beq
{\tilde m}_Q^2(\mu )=\alpha (\mu)^{\ell +1}{\tilde P}_\ell \left(\alpha 
(\mu)L_X\right) ~,
\label{furb}
\eeq
where the dependence on $\luv $ disappears because of the invariance under the
RG, and ${\tilde P}_\ell$ is a function related to the second derivative
of $P_\ell$. 
Therefore it is sufficient to use the RG at the order $\ell =1$ to
obtain the $\alpha^2$ contribution to soft masses.
We conclude
that, while sfermion masses arise from {\it finite} two loop 
diagrams, they can be extracted just by using the one-loop anomalous dimensions.
Soft masses are generated from loop momenta 
of the order of the threshold $M$, but can be reconstructed from
the behaviour of wave-function renormalization far away from threshold 
(RG evolution). 

Let us now derive the well-known results for gauge-mediated soft terms in
the scalar sector by using our technique.
The one-loop RG equation for the wave-function renormalization of a chiral
superfield is
\beq
\frac{d}{dt}\ln Z_Q=\frac{c}{\pi}\alpha ~.
\label{zq}
\eeq
Here $c$ is the quadratic Casimir of the $Q$ gauge representation 
($c=(N^2-1)/(2N)$ for an $SU(N)$ fundamental). The
function $Z_Q(M,\mu)$ is determined by integrating eq.~(\ref{zq})
{}from the ultraviolet scale $\luv$ down to $\mu$ with tree-level
matching at the intermediate threshold $M$. Substituting 
$M\to \sqrt{XX^\dagger}$ we get
\beq
Z_Q(X,X^\dagger,\mu)=Z_Q(\luv )~\left[ \frac{\alpha(\luv )}{\alpha(X)}
\right]^{\frac{2c}{b^\prime}} ~\left[ \frac{\alpha(X )}{\alpha(\mu )}
\right]^{\frac{2c}{b}}~,
\label{zquark}
\eeq
where the $X$ dependence of $\alpha(\mu )$ and $\alpha(X)$ is given by
\beq
\alpha^{-1}(\mu)=16 \pi~{\rm Re} ~S(\mu)=\alpha^{-1}(X)+\frac{b}{4\pi}\ln
\frac{\mu^2}{XX^\dagger}~,
\label{alam}
\eeq
\beq
\alpha^{-1}(X)=16 \pi~
{\rm Re}~ S(X)=\alpha^{-1}(\luv )+\frac{b^\prime}{4\pi}\ln
\frac{XX^\dagger}{\luv^2}~.
\label{ax}
\eeq
Notice that $S(X,\mu)$ is chiral, as required by supersymmetry, 
while $\alpha$, proportional to the real part of $S(X,\mu)$, is not.
Using eq.~(\ref{squak}) and performing the derivatives in $X$ while keeping
the ultraviolet parameters $Z_Q(\luv)$ and $\alpha(\luv)$ fixed,
it is easy to 
obtain the supersymmetry-breaking masses and trilinear terms,
\beq
{\tilde m}_Q^2(\mu)=2c~ \frac{\alpha^2(\mu )}{(4\pi)^2}~N\left[ 
\xi^2+\frac{N}{b}(1-\xi^2)
\right] ~
\left( \frac{F}{M}\right)^2~,
\label{msoft}
\eeq
\beq
A_i(\mu )=\frac{2 c_i}{b}~\frac{\alpha(\mu )}{4\pi}~N(\xi -1)
~\frac{F}{M}~,
\label{asoft}
\eeq
\beq
\xi\equiv \frac{\alpha( M)}{\alpha(\mu)}=\left[
1+\frac{b}{2\pi}\alpha(\mu)
\ln \frac{M}{\mu}\right]^{-1}~.
\eeq
If the superfield $Q$ is charged under different gauge groups, 
eqs.~(\ref{msoft}) and (\ref{asoft}) are generalized by summing over the
different gauge coupling constants. We have recovered the familiar
formulae for the soft terms in theories with gauge-mediated supersymmetry
breaking, including the leading-log effect in the renormalization from
the messenger scale $M$ to the low energy scale $\mu$. In particular, all
$A_i$ vanish at $\mu =M$ ($\xi=1$), but at low-energies acquire 
a renormalization proportional to the gaugino mass.

A few comments are in order at this point.
First of all, our results represent  the lowest order contribution to
soft terms in 
an expansion in powers of $y=F/M^2$. The higher-order terms
correspond to higher-derivative interactions in the gauge and matter
effective kinetic terms. Indeed the expansion parameter is given
by $y=D^2(1/X)$ where $D^2$ is the superspace covariant
derivative squared. These higher-derivative interactions are
genuinely finite effects and cannot be reconstructed with our RG technique.
This is not surprising as our method crucially relies on the approximate
supersymmetry of the low-energy effective Lagrangian and, for $F\sim M^2$,
the messenger spectrum badly violates supersymmetry.
Our technique is however very useful as in a vast class
of interesting models one has $y\ll 1$. Indeed,
as shown by the explicit calculations of refs.~\cite{pom,lui}, 
the asymptotic formulae derived at $y=0$ are 
reached very quickly, so that already
for $F/M^2\sim 0.3$ our approximation is extremely good. This is true
also because
the expansion parameter is actually $y^2$.

Our method as described above crucially relies on the possibility
of substituting unambiguously $M$ with $X$ or $\sqrt {XX^\dagger}$.
For one-loop RG evolution this surely poses no problem. The question
however arises when considering the evolution of $S$ beyond one-loop.
This is because in any practical regularization scheme like DRED, $S$ ceases
being holomorphic in $X$ as soon as two-loop evolution is considered. 
For instance, at two-loops, in the expansion for $\alpha^{-1}
(M,\mu)$, there appear $\ln^n(M/\mu)$ terms with $n>1$. Thus the
analytic continuation $M\to X$ is no longer unambiguous. These higher-order 
effects are absent in the Wilsonian $\beta$ function, for which holomorphy
is explicitly manifest to all orders in perturbation theory \cite{shva}. On the
other hand, for practical purposes, one needs the full gauge
propagator and holomorphy has to be abandoned. The correct prescription,
as discussed in ref.~\cite{hish}, is motivated by the all-order 
formula for $\alpha^{-1}(\mu)$. For instance in an abelian
gauge theory this is given by \cite{svz}
\beq
\alpha^{-1}(M,\mu)=\alpha^{-1}
(\luv)+{b\over 2\pi}\ln{\mu\over M}+{b^\prime\over
2\pi}\ln{M\over \luv}-{1\over 2\pi}\sum_iT_i\ln Z_i(M,\mu) ~,
\label{russians}
\eeq
where $i$ runs over the light matter fields and $T_i$ are the Dynkin
indices. The $\ln Z_i$ can be interpreted as originating
from a non-holomorphic rescaling of the matter fields via the Konishi
anomaly~\cite{kon}. 
The rescaling, albeit non holomorphic, must be supersymmetric:
a chiral superfield must be rescaled with a chiral background wave
function. This is precisely what eq.~(\ref{qp}) does. The right
prescription to promote $\alpha$ to a superfield is then to substitute
$Z$ with ${\cal Z}^\dagger {\cal Z}$, so that the complex version of
eq.~(\ref{russians}) becomes
\beq
S(X,\mu)=S(\luv)+{b\over 32\pi^2}\ln{\mu\over X}+{b^\prime\over
32\pi^2}\ln{X\over \luv}-{1\over 32\pi^2}\sum_iT_i\ln {\cal Z}_i(M,\mu).
\eeq
In the course of this paper we will only be concerned with quantities
for which the one-loop gauge $\beta$ function suffices, so that we will
not get involved into these subtleties.
The only quantities which we study to the next-to-leading accuracy are the
Higgs mass parameters for which the two-loop RG evolution occurs only
in the matter wave-function.

\section{Higgs Mass Parameters at the Next-to-Leading Order}

In the previous section we have computed the supersymmetry-breaking
terms induced by gauge couplings. Neglecting the Yukawa couplings is
usually a good approximation in all practical cases, aside from the Higgs mass
parameters. The large top-quark Yukawa coupling induces a three-loop
contribution to the hypercharge +1 Higgs mass parameter $m_{H}^2\sim \alpha_t
\alpha_s^2/\pi^3\times F^2/M^2$, which is of the same order of magnitude
of the two-loop weak contribution $m_{H}^2\sim \alpha_W^2/\pi^2
\times F^2/M^2$. Moreover, this top-quark Yukawa contribution plays a
key r\^ole in the phenomenology of gauge mediation, since it drives
$m_{H}^2$ negative, triggering electroweak-symmetry breaking.
This contribution is known in the leading-log approximation and we 
will reproduce it here using our method. However, in realistic
gauge-mediation models, the logarithm of the ratio between the messenger
and squark masses $\ln (M/{\tilde m}_Q)\gappeq \ln (2\pi/\alpha_s)$ 
can be as small as 4, and therefore the subleading
constant term is not necessarily
insignificant. The computation of this term requires to go beyond the
leading-log approximation, and it has not been done in previous literature;
we will present it in this section. This example
will illustrate how our method can be simply used to perform computations
which are exceedingly involved if viewed
in terms of Feynman diagrams of component fields.

Let us start with the leading-log calculation. For simplicity we will
consider only top-quark Yukawa and QCD effects, but the inclusion of
electroweak effects is straightforward and gives rise to well-known results.
At the one-loop level, the relevant RG equation for the hypercharge +1 Higgs 
wave-function renormalization is
\beq
\frac{d}{dt}\ln Z_H=-\frac{3}{2\pi}\alpha_t ~.
\label{zh}
\eeq
The RG equation for the top-quark Yukawa coupling, $\alpha_t=h_{top}^2/(4\pi)$,
is 
\beq
\frac{d}{dt}\alpha_t=\frac{\alpha_t}{\pi}~\left( 3\alpha_t-\frac{8}{3}\alpha_s
\right) ~,
\eeq
while the RG equation for the QCD gauge
couplings $\alpha_s$ is given in eq.~(\ref{runalp}), with $b=3$ below the
messenger scale $M$ and $b^\prime =3-N$ above $M$. 
Notice that we are actually calculating the evolution of wave functions
and that, at any energy scale $\mu$,
$\alpha_t=\alpha_t(\luv )Z^{-1}_{Q_L}Z^{-1}_{{\bar U}_R}
Z^{-1}_H$. The use of $\alpha_t$ is a convenient way to describe the evolution
of the product of wave functions $Z_{Q_L}Z_{{\bar U}_R}
Z_H$. 
The solution of 
eq.~(\ref{zh}) is
\beq
Z_H(X,X^\dagger,\mu)=Z_H(\luv )~\left[ 
\frac{\alpha_t(\luv )}{\alpha_t(\mu )}
\right]^{\frac{1}{2}}
~\left[ \frac{\alpha_s(\luv )}{\alpha_s(X)}
\right]^{-\frac{8}{3b^\prime}} ~\left[ \frac{\alpha_s(X )}{\alpha_s(\mu )}
\right]^{-\frac{8}{3b}}~,
\label{zhiggs}
\eeq
where the expressions for $\alpha_s(\mu )$ and $\alpha_s(X )$ are given
in eqs.~(\ref{alam})--(\ref{ax}), and
\beq
\alpha_t(\mu)=\frac{\alpha_t(X)E}{1-\frac{3}{\pi}\alpha_t(X)F}~,
\eeq
\beq
\alpha_t(X)=\frac{\alpha_t(\luv )E^\prime }{1-\frac{3}{\pi}\alpha_t(\luv
)F^\prime}~,
\eeq
\beq
E\equiv
\left[ \frac{\alpha_s(\mu )}{\alpha_s(X)}\right]^{\frac{16}{3b}}~,~~~~~~
F\equiv
\frac{2\pi}{\frac{16}{3}-b}\left[\alpha_s^{-1}(X)-\alpha_s^{-1}(\mu)
E\right] ~,
\eeq
\beq
E^\prime\equiv \left[ \frac{\alpha_s(X )}{\alpha_s(\luv )}\right]^{
\frac{16}{3b^\prime}}~,~~~~~~
F^\prime \equiv \frac{2\pi}{\frac{16}{3}-b^\prime}\left[\alpha_s^{-1}(\luv )-
\alpha_s^{-1}(X)
E^\prime \right] ~.
\eeq

Using the general formula in eq.~(\ref{squak}), we can express the
top-quark Yukawa contribution to $m_H^2$ as
\beq
m_H^2(\mu)=\frac{\alpha_t(\mu)}{\pi^2}~N~(1-\xi)\left\{
N(1-\xi )\left[ \frac{\alpha_t(\mu )}{8}(1-I)^2-\frac{\alpha_s(\mu)}{9}
\right] -\frac{\alpha_s(\mu)}{3}\xi I\right\}~\left(
\frac{F}{M}\right)^2
~,
\label{mh}
\eeq
\beq
\xi\equiv \frac{\alpha_s( M)}{\alpha_s(\mu)}=\left[
1+\frac{3}{2\pi}\alpha_s(\mu)
\ln \frac{M}{\mu}\right]^{-1}~,
\eeq
\beq
I\equiv \frac{9}{7}~\frac{(1-\xi^{\frac{7}{9}})\xi}{(1-\xi)}~.
\eeq
If the logarithm $\ln (M/\mu )$ is not too large, it is convenient
to expand eq.~(\ref{mh}) and obtain the three-loop contribution in the
leading-log approximation:
\beq
m_H^2(\mu)=-\frac{\alpha_t(\mu)\alpha_s^2(\mu)}{2\pi^3}~N~
\left( \ln \frac{M}{\mu}\right) ~\left(
\frac{F}{M}\right)^2~.
\label{trenta}
\eeq
This is the known negative contribution to the Higgs mass squared parameter
which leads to electroweak-symmetry breaking. In the following we will
compute the three-loop $\alpha_t \alpha_s^2$ term with no logs. 

Before proceeding to the next-to-leading order calculation, we remark
that, for gauge-mediated models in which the Higgs mixing mass
$\mu_H$ is a hard parameter\footnote{In order not to generate confusion 
between the renormalization scale and the Higgs mixing mass parameter,
we have chosen to denote the latter by $\mu_H$, instead of the usual symbol
$\mu$.} ({\it i.e.} $\mu_H$ is not generated by loops
involving messenger fields), then eq.~(\ref{zhiggs}) also provides 
the expression of the bilinear parameter $B$ at low energies. From 
eqs.~(\ref{pot}) and (\ref{aa}), we find
\beq
B(\mu)=\frac{\alpha_t(\mu)}{4\pi}~N~(1-\xi)(1-I)~\frac{F}{M}~,
\label{b}
\eeq
which, expanded to the two-loop level, becomes
\beq
B(\mu)=\frac{\alpha_t(\mu)\alpha_s^2(\mu)}{2\pi^3}~N
~\left( \ln^2 \frac{M}{\mu}\right) ~\frac{F}{M}~.
\label{bex}
\eeq
If the mechanism responsible for generating the $\mu_H$ term also gives
a contribution to $B$, this should be added to eqs.~(\ref{b}) and
(\ref{bex}).

Let us now turn to the next-to-leading order calculation. We are interested
in the subleading 
contribution to the term ${\cal O}(\alpha_t\alpha_s^2)$,
{\it i.e.} the term suppressed by $1/\ln (M/\mu)$ with respect
to eqs.~(\ref{trenta}) and (\ref{bex}). For this purpose
it
is sufficient to retain only the ${\cal O}(\alpha_t\alpha_s)$ term in the 
RG equation for $Z_H$, the ${\cal O}(\alpha_t\alpha_s^2)$ term in the
equation for $\alpha_t$,
and use the RG equation for $\alpha_s$ at the leading order, eq.~(\ref{runalp}).
In the $\overline{DR}$ scheme, the relevant RG equations are~\cite{twoloop}
\beq
\frac{d}{dt}\ln Z_H=-\frac{3}{2\pi}\alpha_t-\frac{2}{\pi^2}\alpha_t\alpha_s ~,
\label{znex}
\eeq
\beq
\frac{d}{dt}\alpha_t=\frac{\alpha_t}{\pi}~\left( 3\alpha_t-\frac{8}{3}\alpha_s
\right)-\frac{2d}{\pi^2}\alpha_t\alpha_s^2 ~,
\label{runnex}
\eeq
In eq.~(\ref{runnex}), $d=1/9$ below the messenger scale $M$ and 
$d=(1-3N)/9$ above $M$. Expanding at the three-loop level the solution to 
the differential equation (\ref{znex}), we obtain
\beq
\ln \frac{Z_H(X,X^\dagger, \mu)}{Z_H(\luv )}=\frac{\alpha_t(\mu)
\alpha_s^2(\mu)}{(2\pi)^3}~N~\left[ \frac{1}{3}\ln^3\left(
\frac{XX^\dagger}{\mu^2}\right) +2\ln^2\left(
\frac{XX^\dagger}{\mu^2}\right) \right] +f[ \ln (\luv /\mu )] ~.
\label{next}
\eeq
Here $f$ is a function of $\ln (\luv /\mu) $ and it is independent of $X$.
The log$^2$ term in eq.~(\ref{next}) represents the subleading correction.
Calculation of the linear-log term (which contributes to $B$, but not to
$m_H^2$) would require knowledge of the next order in 
perturbation theory for the $\beta$ functions and the anomalous dimensions.
{}From eqs.~(\ref{squak}) and (\ref{aa}), we obtain
\beq
m_H^2(\mu)=-\frac{\alpha_t(\mu)\alpha_s^2(\mu)}{2\pi^3}~N~
\left(\ln \frac{M}{\mu}+1\right) ~\left(
\frac{F}{M}\right)^2~.
\label{fin1}
\eeq
\beq
B(\mu)=\frac{\alpha_t(\mu)\alpha_s^2(\mu)}{2\pi^3}~N
~\left(\ln^2 \frac{M}{\mu}+2\ln \frac{M}{\mu}\right)~\frac{F}{M}~.
\label{fin2}
\eeq

To complete the next-to-leading order calculation we have to include the
matching conditions at the ultraviolet scale $M$ and the infrared scale
$\mu$. 
Notice that even though we only need to evolve $\alpha_s$ at one-loop,
it is crucial to correctly match $\alpha_s$ at the next-to-leading
order in the chosen scheme.
In the $\overline{DR}$ scheme, the matching of the running
gauge coupling constant for thresholds of heavy scalars and fermions
has to be done at a scale equal to the mass of the heavy particles 
\cite{weha}.
Therefore in eqs.~(\ref{fin1})--(\ref{fin2}) 
we identify $M$ with the physical mass of the messenger fields.
To determine the
infrared matching, one has to include the one-loop effective potential
which cancels the $\mu$ dependence at the appropriate order in
perturbation theory. This can be seen explicitly by expanding the
effective potential in powers of ratios between the Higgs fields
$H$, $\bar H$ and the supersymmetry-breaking stop mass ${\tilde m}_t$, see
eq.~(\ref{msoft}). 
This is a good approximation since, in realistic models, the stop turns out
to be rather heavy.
In the $\overline{DR}$ scheme, we find
\bea
&&V_{1-loop}=\frac{1}{64\pi^2}{\rm STr} {\cal M}^4 \left(\ln\frac{{\cal M}^2}
{\mu^2}-\frac{3}{2}\right)\simeq \nonumber \\
&&\simeq \frac{3\alpha_t(\mu)}{2\pi}
\left[ |A_t(\mu)H-\mu_H {\bar H}^\dagger |^2\ln \frac{{\tilde m}_t(\mu)
}{\mu}
+|H|^2{\tilde m}_t(\mu)^2\left(2\ln \frac{{\tilde m}_t(\mu)
}{\mu} -1\right)\right]
~.
\eea
Here $A_t$ is the coefficient of the supersymmetry-breaking trilinear
interaction for the stop, and $\mu_H$ is the supersymmetric Higgs
mixing mass.
Therefore $V_{1-loop}$ gives an effective contribution to the soft-breaking
parameters
\bea
&&\delta m_H^2 = \frac{3\alpha_t(\mu)}{2\pi}\left[ A_t^2(\mu)
\ln \frac{{\tilde m}_t(\mu)
}{\mu}+{\tilde m}_t^2(\mu)
\left(2\ln \frac{{\tilde m}_t(\mu)
}{\mu} -1\right)\right]
\simeq \nonumber \\
&&\simeq \frac{\alpha_t(\mu)\alpha_s^2(\mu)}
{2\pi^3}~N~\left(\ln 
\frac{{\tilde m}_t(\mu)
}{\mu} -\frac{1}{2}\right) ~\left( \frac{F}{M}\right)^2 ~,
\label{e1}
\eea
\beq
\delta B = -\frac{3\alpha_t(\mu)}{2\pi}A_t(\mu)
\ln \frac{{\tilde m}_t(\mu)
}{\mu} \simeq\frac{\alpha_t(\mu) \alpha_s^2(\mu)}
{\pi^3}~N~
\ln \frac{{\tilde m}_t(\mu)
}{\mu} \ln \frac{M}{\mu}~\frac{F}{M}~.
\label{e2}
\eeq
The right-hand sides of eqs.~(\ref{e1})--(\ref{e2}) are obtained by
retaining only the genuine three-loop effects. 
These contributions can be reabsorbed in the definition of effective
supersymmetry-breaking parameters.
By adding eqs.~(\ref{e1})--(\ref{e2})  to
eqs.~(\ref{fin1})--(\ref{fin2}) 
we obtain that the effective values of the
supersymmetry-breaking parameters are
\beq
m_{H{\rm eff}}^2=-\frac{\alpha_t({\tilde m}_t)\alpha_s^2({\tilde m}_t)
}{2\pi^3}~N~
\left(\ln \frac{M}{{\tilde m}_t}+\frac{3}{2}\right) ~\left(
\frac{F}{M}\right)^2~.
\label{ffin1}
\eeq
\beq
B_{\rm eff}=\frac{\alpha_t({\tilde m}_t)\alpha_s^2({\tilde m}_t)}{2\pi^3}~N
~\left(\ln^2 \frac{M}{{\tilde m}_t}+2\ln \frac{M}{{\tilde m}_t}
\right)~\frac{F}{M}~.
\label{ffin2}
\eeq
The result in eq.~(\ref{ffin2}) agrees with the explicit two-loop calculation
of the component-field Feynman diagrams presented 
in eqs.~(27) and (29) of
ref.~\cite{rat}.
The next-to-leading correction to $m_H^2$, which has been calculated
here for the first time, can be as large as 30\%. This
modifies by the same amount the extraction of the $\mu_H^2$ parameter from
the electroweak-breaking condition and consequently the physical mass
spectrum of charginos and neutralinos.

\section{Superpotential Couplings between Messengers and Matter}

Models with gauge-mediated supersymmetry breaking usually assume that
messengers and ordinary matter do not couple directly in the 
superpotential. This is to avoid any new source of flavour breaking,
which is, after all, the primary motivation for these models. However,
$SU(3)\times SU(2)\times U(1)$ gauge invariance alone does not forbid
such couplings. Actually superpotential interactions involving 
messenger and Higgs
superfields were considered in ref.~\cite{noi}, while 
interactions between messengers
and quarks or leptons were studied in ref.~\cite{nir}. 
It was shown \cite{noi,nir} that these couplings
do not induce at the leading order in $F/M$ one-loop 
supersymmetry-breaking masses for the scalar components
of the matter superfields. This property was explained in ref.~\cite{soft}
using superfield language. In this section we will give the general
expression for supersymmetry-breaking terms induced by messenger-matter
couplings, including the two-loop soft masses for the scalar components
of chiral superfields.

We will consider a coupling in the superpotential 
between one matter and two messenger superfields, $\lambda Q \Phi_1 \Phi_2$,
or 
between two matter and one messenger superfields, $\lambda Q_1 Q_2 \Phi$. 
The formulae we present are the same in both cases. For simplicity we
assume that all superfields are charged under a simple gauge group.
The generalization to gauge groups made of
direct products of simple groups is straightforward.
The RG equations for the wave-function renormalization of the 
matter superfield $Q_i$ is
\beq
\frac{d}{dt}\ln Z_{Q_i}=\frac{1}{\pi}\left( c_i\alpha-\frac{d_i}{2}
\alpha_\lambda
\right) ~,
\label{zhh}
\eeq
with $\alpha_\lambda
\equiv \lambda^2/(4\pi)$. Here $c_i$ is the quadratic Casimir of the
$Q_i$ gauge representation and $d_i$ is the number of fields circulating
in the Yukawa loop. For instance, if the $\lambda$ interaction involves
a fundamental, an anti-fundamental, and a singlet of $SU(N)$, then
$d_i=1$ or $N$ if $Q_i$ is a fundamental or a singlet, respectively.
The RG equation for the gauge coupling constant is given in eq.~(\ref{runalp})
and the one for $\alpha_\lambda$ is
\beq
\frac{d}{dt}\alpha_\lambda=\frac{\alpha_\lambda}{\pi}~\left( \frac{D}{2}
\alpha_\lambda
-C\alpha \right) ~.
\eeq
Here $C=\sum_i c_i$, $D=\sum_i d_i$ with the sum extended to all
superfields participating to the $\lambda$ interaction.
The solution to eq.~(\ref{zhh}) is
\beq
Z_{Q_i}(X,X^\dagger,\mu)=Z_{Q_i}(\luv )Z_{GM}(X,X^\dagger,\mu)
Z_{\lambda}(X,X^\dagger,\mu)
\eeq
\beq
Z_{GM}(X,X^\dagger,\mu)\equiv\left[ 
\frac{\alpha(\luv )}{\alpha(X)}
\right]^{\frac{2c_i}{b^\prime}} ~\left[ \frac{\alpha(X )}{\alpha(\mu )}
\right]^{\frac{2c_i}{b}}~,
\label{zgm}
\eeq
\beq
Z_{\lambda}(X,X^\dagger,\mu)\equiv\left[ 
\frac{\alpha_\lambda(\luv )}{\alpha_\lambda(X)}
\right]^{\frac{d_i}{D}} ~\left[ \frac{\alpha(\luv )}{\alpha(X )}
\right]^{-\frac{2Cd_i}{Db^\prime}}~,
\label{zl}
\eeq
\beq
\alpha_\lambda(X)=\frac{\alpha_\lambda(\luv )E^\prime}{1-\frac{D}{2\pi}
\alpha_\lambda
(\luv )F^\prime}~,
\eeq
\beq
E^\prime \equiv \left[ \frac{\alpha(X )}{\alpha(\luv )}\right]^{
\frac{2C}{b^\prime}}~,~~~~~~
F^\prime \equiv \frac{2\pi}{2C-b^\prime}\left[\alpha^{-1}(\luv )-
\alpha^{-1}(X)
E^\prime \right] ~.
\eeq
The factor $Z_{GM}$ is the usual contribution from gauge mediation, while
$Z_{\lambda}$ is the new contribution from the $\lambda$ interaction.
Notice that $Z_{\lambda}$ is independent of $\mu$ (for $\mu < M$)
since the $\lambda$ interaction is not present in the effective theory
with messenger fields integrated out.
The soft terms (which originate from differentiating the 
logarithm of $Z$) are given by the sum of the gauge-mediated 
contribution, given in eqs.~(\ref{msoft})--(\ref{asoft}),
and a genuine new contribution given
by
\beq
\delta {\tilde m}_{Q_i}^2(\mu )=\frac{d_i}{8\pi^2}\alpha_\lambda(M)
\left[\frac{D}{2}\alpha_\lambda
(M)-C\alpha (M)\right] \left ({F\over M}\right)^2~,
\label{lamas}
\eeq
\beq
\delta A_i(\mu)=-\frac{ d_i}{4\pi}\alpha_\lambda(M){F\over M}~.
\label{laaa}
\eeq

As expected, supersymmetry-breaking masses vanish at one loop at the leading
order in $F/M^2$. The one-loop contribution proportional to $(\alpha_\lambda
/4\pi)(F^4/M^6)$ computed in ref.~\cite{nir} is smaller than 
the two-loop expression shown in eq.~(\ref{lamas}), as long as
$F/M^2\lappeq \sqrt{\alpha_s/(4\pi)}$. In contrast to the
case of gauge mediation, $A$ terms are generated at one loop, see 
eq.~(\ref{laaa}). Therefore, if superpotential matter-messenger interactions
exist, their contribution to supersymmetry-breaking parameters can 
substantially modify the physical spectrum of new particles.

The superpotential $\lambda$ interactions can find an interesting application
in the generation of the Higgs-mixing $\mu_H$ parameter. It is known \cite{noi}
that gauge-mediation models have difficulties in generating dynamically
the $\mu_H$ term. In particular, the introduction of a singlet superfield $N$
with superpotential couplings $W=NH{\bar H}+N^3$ does not easily
solve the problem 
because, at the messenger scale $M$, all supersymmetry-breaking terms 
involving $N$ vanish at the leading order. A negative mass squared for
the $N$ scalar component and trilinear couplings are generated by the
RG evolution to the weak scale, but these parameters turn out to be too small
to insure a correct symmetry-breaking pattern and an acceptable mass spectrum.
A possible solution invoked in ref.~\cite{din} is to introduce new light
particles which increase the renormalization effects.
 
The results shown in this section open a different possibility. Suppose
$N$ interacts with the messenger fields $\Phi$, $\bar \Phi$
with a superpotential coupling
\beq
W=\lambda N {\bar \Phi}\Phi ~.
\label{npp}
\eeq
Now the scalar component of $N$ gets a two-loop
negative mass squared proportional to $\alpha_\lambda \alpha_s$, see 
eq.~(\ref{lamas}), where 
$\alpha_s$ is the QCD coupling constant. Moreover supersymmetry-breaking
trilinear terms involving the scalar field $N$ are generated at one loop.
The electroweak-breaking conditions can now be much more
easily satisfied. Notice
however that the superfield $N$ has the same quantum numbers as the Goldstino
superfield $X$, and therefore the wave-function renormalization mixes 
these two superfields generating a term $\int d^4\theta X^\dagger N$. After
supersymmetry breaking this term leads to a tadpole for $F_N$ larger than the
electroweak scale. This can be avoided by extending the messenger sector
in such a way that $N$ and $X$ carry different quantum numbers under some
new symmetry. A simple example is the case of two messenger flavours with
superpotential
\beq
W=X({\bar \Phi}_1 \Phi_1 +{\bar \Phi}_2 \Phi_2 ) +N{\bar \Phi}_1 \Phi_2~.
\eeq 
The theory has a discrete $Z_3$ symmetry under which all chiral supefields
(including Higgs and matter) have charge 1/3, but for $\Phi_1$ and
${\bar \Phi}_2$ which have charges -1/3 and for $X$ which is neutral.
This symmetry,
broken only at the weak scale, distinguishes between $X$ and $N$.
Finally we just remark that a different possibility is
using an interaction of the form
$\lambda NH{\bar \Phi}$ instead of eq.~(\ref{npp}). We have not investigated
if this can lead to a successful mechanism of electroweak breaking.

\section{Gauge Messengers}

So far we have considered the most familiar case in which the messenger
particles belong to chiral superfields. However it is also possible that
gauge supermultiplets behave as messengers. We are envisaging a situation
in which the supersymmetry-breaking VEV is also responsible for spontaneous
breaking of some gauge symmetry containing the Standard Model (SM) 
as a subgroup. 
The vector bosons corresponding to the broken generators, together with
their supersymmetric partners,
receive masses proportional to $M$. However supersymmetry-breaking
effects proportional to $F$ split the gauge supermultiplets at tree level
and consequently soft terms for observable fields are generated 
by quantum effects.
In this section we will compute these terms. We will show that
our method enormously simplifies a calculation which is complicated if
performed by evaluating component-field Feynman diagrams. But this is not
merely an exercise to show the power of our method. We will find that the
contribution from gauge messengers, which have never been calculated before,
are of the same order as those from
chiral messengers and therefore can significantly
modify the phenomenology of certain classes of models. 
For instance, this happens
in models based on Witten's inverse 
hierarchy~\cite{wit} or in more recent proposals that combine the
inverse hierarchy with dynamical supersymmetry breaking~\cite{mur,soft}.

Let us first consider the case in which $\langle X\rangle$ spontaneously
breaks the gauge group $G\to H$, in such a way that the gauge coupling
constant is continuous at the threshold. 
The calculation of the soft term generated
by gauge interactions is completely analogous to the one presented in sect.~2
for chiral messengers, and gives
\beq
{\tilde M}_g (\mu)=\frac{\alpha(\mu)}{4\pi}~N~\frac{F}{M} ~,
\label{gaugaums}
\eeq
\beq
{\tilde m}_Q^2(\mu)=2c~ \frac{\alpha^2(\mu )}{(4\pi)^2}~N\left\{ 
\xi^2 \left[ 1+r\left( \frac{N}{b}-1\right) \right]
+\frac{N}{b}(1-\xi^2)
\right\} ~
\left( \frac{F}{M}\right)^2~,
\label{msoftms}
\eeq
\beq
A_i(\mu )=\frac{2 c_i}{b}~\frac{\alpha(\mu )}{4\pi}~N(\xi  -1-r\xi )
~\frac{F}{M}~.
\label{asoftms}
\eeq
Here $N$ is the generalization of the messenger index, defined as
the difference between the low-energy ($H$) and high-energy ($G$)
$\beta$-function coefficients:
\beq
N=b-b^\prime = N_f -2(C_G-C_H)~.
\label{vecmes}
\eeq
For chiral messengers, this is just the
number of flavours of heavy multiplets $N_f$. In the case of gauge
messengers, one has to add the contribution of heavy gauge bosons, equal
to $-3(C_G-C_H)$, and the contribution of the would-be Goldstone bosons,
equal to $(C_G-C_H)$. Here $C_G$ is the quadratic Casimir of the 
adjoint representation of $G$,
equal to $N$ for an $SU(N)$ group.
Therefore, the pure gauge-messenger effect is to
reduce the value of $N$, and allow also negative values of the total $N$. 
In eqs.~(\ref{msoftms})--(\ref{asoftms}), the coefficient $r$ is defined as
\beq
r=\frac{\frac{c^\prime}{c}-1}{\frac{b^\prime}{b}-1}~,
\eeq
where $c$ and $c^\prime$ are the quadratic Casimirs of the matter gauge
representations of the groups $H$ and $G$, respectively. Notice that the
trilinear terms $A_i$ are non-vanishing even at the messenger scale ($\xi
=1$). Moreover, the scalar squared masses ${\tilde m}_Q^2$ have a negative
boundary condition at the messenger
scale whenever $b^\prime (2-c^\prime /c)>b$.

Before discussing phenomenological applications, we believe it is useful to 
understand our results in terms of component fields. 
To follow the same procedure used in the case of
chiral messengers, we first have to determine the tree-level mass
splittings inside the gauge messenger supermultiplets, and then study how this
is tranferred to the observable sector at the quantum level. The
messenger mass spectrum can be derived from the kinetic terms
\beq
{\cal L}=
\int d^4 \theta X^\dagger e^V X +
\left(
\int d^2 \theta \frac{1}{4g^2}W^{a\alpha}W^a_\alpha + {\rm h.c.}\right) ~.
\label{kine}
\eeq
Instead of the usual Wess-Zumino gauge, it is more convenient to use
a unitary gauge for the vector superfield, which eliminates the
would-be Goldstone boson chiral multiplets from the superfield $X$.
In this gauge, the vector superfield $V$ contains a vector boson $v_\mu$,
two Weyl fermions $\lambda$ and $\chi$, a real scalar $C$, one real and
one complex auxiliary fields $D$ and $N$ (we follow the conventions of
ref.~\cite{wess} and suppress the gauge index $a$). 
In order to have canonical kinetic terms and canonical
field dimensions, we do the rescaling $(\lambda ,v_\mu ,D)\to g
(\lambda ,v_\mu ,D)$ and $(C, \chi , N)\to M (C, \chi , N)$, where $g$
is the gauge coupling constant and $M$ is the VEV of the scalar component
of the Goldstino superfield. Expanding eq.~(\ref{kine}) in field bilinears,
we find the usual kinetic terms for all propagating fields and the 
mass terms
\bea
{\cal L}&=&\frac{1}{2}(gM)^2v^\mu v_\mu +\frac{1}{2}D^2
+\frac{1}{2}N^\dagger N+gMCD-gM(\chi \lambda +{\rm h.c.})+\nonumber \\
&+&\frac{F}{M}\left(
iCN-\frac{1}{2}\chi \chi +{\rm h.c.}\right) +\frac{F^2}{M^2}C^2 ~.
\eea
Eliminating the auxiliary fields $D$ and $N$, we obtain
\beq
{\cal L}=\frac{1}{2}(gM)^2v^\mu v_\mu -\frac{1}{2}\left( g^2M^2
+2\frac{F^2}{M^2}\right) C^2-\frac{1}{2}\left( \psi^T {\cal M}_\psi \psi
+{\rm h.c.}\right) ~,
\eeq
\beq
\psi \equiv \pmatrix{\lambda \cr \chi} ~~~~~{\cal M}_\psi \equiv
\pmatrix{0& gM\cr gM & \frac{F}{M}} ~.
\eeq
Therefore the gauge multiplet contains a vector boson with mass $gM$, two
Weyl fermion with masses $[\sqrt{4g^2M^2+(F/M)^2}\pm (F/M)]/2$, and one real
scalar with mass squared $g^2M^2+2(F/M)^2$. Notice that the 
supersymmetry-breaking VEV $F$ spoils the usual superHiggs mass relation,
but preserves the condition of vanishing supertrace.

Our result for the gauge-messenger contribution to gaugino
masses, see eq.~(\ref{vecmes}), can also be extracted from the result
of an explicit component calculation of the GUT threshold corrections
to gaugino masses~\cite{goto}. For illustrative purposes, we want to
rederive the same result from a component calculation in a simple model.
Let us consider an $SU(2)$ gauge theory which is broken, in the supersymmetric
limit, to $U(1)$ by the VEV of a triplet chiral superfield $X^i$ ($i=1,2,3$),
along the direction $\langle X_{1,2}\rangle =0$, $\langle X_3\rangle \ne 0$.
We include the effect of supersymmetry breaking by introducing a superpotential
interaction of $X$ with an external source $F$
\beq
W=F\sqrt{\vec{X}^2}=F\langle X_3 \rangle \left( 1+\frac{X_+X_-}{\langle X_3
\rangle ^2}+...
\right) ~.
\label{supp}
\eeq
The interaction in eq.~(\ref{supp}) describes a 
gauge-invariant implementation
of the condition $\langle F_{X_3}\rangle 
=F$, $\langle F_{X_{1,2}}\rangle =0$. Since we are interested in the mass
of the light $U(1)$ gaugino (the ``photino") at one loop, we just need to know 
the interaction Lagrangian for the charged fields. For this reason we have 
expanded eq.~(\ref{supp}) around the X VEV, and retained only the leading terms 
involving $X_\pm =(X_1\pm i X_2)/\sqrt{2}$. In the Feynman-'t Hooft gauge
and working at the first order in $F$,
the scalar components of $X_+$ and $X_-$ are mass degenerate and therefore
there is no physical mixing angle between the two states. On the other
hand, the supersymmetry-breaking mass for the fermionic components of
$X_+$ and $X_-$, see eq.~(\ref{supp}), flips $X_+$ into $X_-$. Thus, at
the first order in $F$ (and in the Feynman-'t Hooft gauge), 
there is no contribution to the ``photino" mass
{}from loops involving scalar and fermionic components of the ``higgsinos"
$X_\pm$. All we need
to do is to compute a loop diagram with charged gauge bosons and gauginos.
This gives a ``photino mass"
\beq
{\tilde M}=-\frac{\alpha}{\pi}~\frac{F}{\langle X_3\rangle}~.
\eeq
This result agrees with eq.~(\ref{vecmes}). The high-energy $SU(2)$ theory
with the chiral triplet $X$ has $b^\prime =6-2=4$, while $b=0$ since there
are no light charged chiral superfields. The two-loop calculation of the 
supersymmetry-breaking scalar masses from gauge messengers
is of course much more involved.

We want to apply now our results to
the scenarios proposed in ref.~\cite{mur,soft}. This requires to study the
case in which $\langle X\rangle$ breaks the gauge
group $G\times H \to H^\prime$, 
where $H^\prime$ is the same group as $H$, but it is non-trivially 
embedded in $G\times H$.
The coupling constant $\alpha$ of
the low-energy gauge group $H^\prime$ is
related to the coupling constants of the
high-energy groups $G$ and $H$ by $\alpha^{-1}=\alpha_G^{-1}+ \alpha_H^{-1}$.
The case of an arbitrary mixing angle between gauge coupling constant 
is a trivial generalization
of the formulae we present below. We also assume that ordinary matter
is charged under $H$ and $H^\prime$, but not under $G$.

The wave-function renormalizations for the gauge and matter
superfields are, respectively
\beq
S(X,\mu)=S(\luv )+\frac{b_H+b_G}{32\pi^2}\ln
\frac{X}{\luv}+\frac{b}{32\pi^2}\ln
\frac{\mu}{X}~,
\label{sxb}
\eeq
\beq
Z_Q(X,X^\dagger,\mu)=Z_Q(\luv )~\left[ \frac{\alpha_H(\luv )}{\alpha_H(X)}
\right]^{\frac{2c}{b_H}} ~\left[ \frac{\alpha(X )}{\alpha(\mu )}
\right]^{\frac{2c}{b}}~,
\label{zquarkb}
\eeq
\beq
\alpha^{-1}(\mu)=\alpha^{-1}(X)+\frac{b}{4\pi}\ln
\frac{\mu^2}{XX^\dagger}~,
\label{alamb}
\eeq
\beq
\alpha^{-1}(X)=\alpha_G^{-1}(\luv )+\alpha_H^{-1}(\luv )+
\frac{(b_H+b_G)}{4\pi}\ln
\frac{XX^\dagger}{\luv^2}~,
\label{axb}
\eeq
\beq
\alpha_H^{-1}(X)=\alpha_H^{-1}(\luv )+
\frac{b_H}{4\pi}\ln
\frac{XX^\dagger}{\luv^2}~.
\eeq
Here $b_G$, $b_H$, and $b$ are the $\beta$-function coefficients of the
high-energy groups $G,H$ and the low-energy group $H^\prime$, respectively.
{}From the general equations (\ref{gaugino}), (\ref{squak}), and 
(\ref{aa}) we obtain
the expressions for the supersymmetry-breaking gaugino and scalar masses
\beq
{\tilde M}_g (\mu)=\frac{\alpha(\mu)}{4\pi}~(b-b_H-b_G)~\frac{F}{M}~,
\label{glub}
\eeq
\beq
{\tilde m}_Q^2=2c~ \frac{\alpha^2(\mu )}{(4\pi)^2}~
\left\{ 
\left[ b+(R^2-2)b_H-2b_G\right] \xi^2 +\frac{(b-b_H-b_G)^2}{b}(1-\xi^2)
\right\} ~
\left( \frac{F}{M}\right)^2~.
\label{msoftb}
\eeq
\beq
A_i(\mu )=\frac{2 c_i}{b}~\frac{\alpha(\mu )}{4\pi}~
\left[(bR-b_H-b_G)\xi -(b-b_H-b_G)\right]
~\frac{F}{M}~,
\label{asoftb}
\eeq
\beq
\xi\equiv \frac{\alpha( M)}{\alpha(\mu)}=\left[
1-\frac{b}{2\pi}\alpha(\mu)
\ln \frac{M}{\mu}\right]^{-1}~,
\eeq
\beq
R\equiv \frac{\alpha_H(M)}{\alpha (M)}=1+\frac{\alpha_H(M)}{\alpha_G(M)}~.
\eeq
Gauge messengers
give a new contribution to scalar masses at $\mu =M$ ($\xi=1$) proportional to
$-2b_G$, see eq.~(\ref{msoftb}). For an asymptotically-free group $G$, these
contributions are negative and can destabilize squarks and sleptons. The
RG evolution proportional to the gaugino mass squared helps to restore
positivity, but there are strong constraints on the group $G$, as we will
show below. It is also interesting to notice that the trilinear $A$ terms
are generated even without running ($\xi =1$), since $R>1$, see 
eq.~(\ref{asoftb}).

These formulae can now be
directly applied to one of the examples discussed
in ref.~\cite{soft}: the ``$SU(5)^3$ model". We have to identify $G$ with
$SU(5)$, $H=H^\prime$ with
$SU(3)\times SU(2)\times U(1)$, $b$ with the SM coefficients, and $b_H=
b-5$, $b_G=10$. It is easy to see from eq.~(\ref{msoftb}) that the right-handed
sleptons turn out to have negative squared masses. This disease could
be cured by adding matter charged under $G$ in order to reduce $b_G$,
source of the negative contribution. Making the group $G$ less
asymptotically free can be dangerous, since this can destabilize the minimum
of $X$ along the flat direction found in ref.~\cite{soft}. 
This is a general difficulty of models with gauge messengers. 
Nevertheless, the other models presented in ref.~\cite{soft}
are viable, since $\langle X\rangle$ does not spontaneously break a gauge
group which contains the SM.

Another example of a model with gauge messengers has been proposed by
Murayama~\cite{mur}. Although the Goldstino resides in a single field
($\Sigma$ in the notation of ref.~\cite{mur}), the model has three
different mass thresholds, because some fields ($S$ and $\phi$) acquire
their masses from Planck-suppressed higher-dimensional operators. The
generalization of the previous equations to this case is straightforward.
Identifying $H=H^\prime$ with the SM group and $G$ with $SU(5)$, we 
obtain the following expressions for gaugino and scalar masses:
\beq
{\tilde M}_g (\mu)=\frac{\alpha(\mu)}{4\pi}~(4b-b_\phi-2b_S-b_H-b_G)
~\frac{F}{M}~,
\label{glubm}
\eeq
\bea
{\tilde m}_Q^2&=&2c~ \frac{\alpha^2(\mu )}{(4\pi)^2}~
\left[ R^2b_H\frac{\alpha^2(M_Q)}{\alpha^2(\mu)}
-\frac{(b_H+b_G)^2}{b_S}\frac{\alpha^2(M_Q)}{\alpha^2(\mu)}
+\right.\nonumber \\
&+&\left( \frac{1}{b_S}-\frac{1}{b_\phi}\right)(b_H+b_G+2b_S)^2
\frac{\alpha^2(M_S)}{\alpha^2(\mu)}+
 \left( \frac{1}{b_\phi}-\frac{1}{b}\right)(b_H+b_G+2b_S+b_\phi)^2
\frac{\alpha^2(M_\phi)}{\alpha^2(\mu)}+ \nonumber \\
&+&\left.
\frac{1}{b}(b_H+b_G+2b_S+b_\phi-4b)^2 \right] \left( \frac{F}{M}\right)^2 ~,
\label{massm}
\eea
\beq
R\equiv \frac{\alpha_H(M_Q)}{\alpha 
(M_Q)}=1+\frac{\alpha_H(M_Q)}{\alpha_G(M_Q)}~.
\eeq
The three mass thresholds are given by $M_Q=\lambda v/\sqrt{5}$,
$M_S=(v/\sqrt{5})^3/M_{Pl}^2$, $M_\phi=(v/\sqrt{5})^4/M_{Pl}^3$, where 
$M_{Pl}$ is the reduced Planck mass, and $v$ is the symmetry-breaking VEV
lying somewhere between $3\times 10^{14}$ GeV and the GUT scale. The
$\beta$-function coefficients $b$ are those of the SM, and 
$b_\phi =b-2$, $b_S=b-5$, $b_H=b-9$, $b_G=6$. 

The coefficient $(4b-b_\phi-2b_S-b_H-b_G)$ of the gaugino mass, see
eq.~(\ref{glubm}) is equal to 15. Had we ignored the gauge messenger
contribution, this would be equal to 25. The gauge messenger effect
is also important for the scalar masses, see eq.~(\ref{massm}). We find
that at least the right-handed slepton has negative mass squared, unless
$v$ coincides with the GUT scale and $\alpha_G$ is large ($R$ close to 1).
Of course in this case one is sensitive to GUT physics, and one should
investigate specific unified models.

The model of ref.~\cite{mur} is less problematic than the ``SU(5) model"
of ref.~\cite{soft}. This is because of the larger matter content and 
because the higher-dimensional couplings between messengers and Goldstino
superfield enhance the effective supersymmetry-breaking scale $F/M$. 
However a complete study of the minimization conditions is still lacking.
At any rate, even in the model of ref.~\cite{mur}, right handed sleptons
have negative squared masses, unless $v$ becomes equal to the GUT scale.

In conclusion, gauge messengers drastically change the usual mass relations
obtained with chiral messengers. However, in the models presented in the 
literature, the new contributions to scalar mass squared from gauge messengers
turn out to be negative and spoil the viability of the models. It would be
interesting to construct realistic theories with gauge messengers.

\section{Conclusions}

We have described a new simple method to derive the supersymmetry-breaking
terms for the observable sector mediated by renormalizable perturbative
interactions. We are dealing with theories in which the relevant mass
scales are set by the VEV of the Goldstino chiral superfield $X$,
$\langle X \rangle =M+\theta^2 F$. Some fields, called the messengers,
receive masses of order $M$ from their tree-level couplings to $X$. 
Assuming $F\ll M^2$, we can use the non-renormalization theorems of
supersymmetry, and obtain the $M$ functional dependence of the wave-function
renormalizations from the RG evolution through the threshold $M$. Then
we perform the appropriate replacement of $M$ with the chiral superfield $X$,
dictated either by holomorphy or by chiral reparametrization. Finally, 
by replacing $X$ with its VEV $\langle X \rangle =M+\theta^2 F$, we obtain
all supersymmetry-breaking effects, at the leading order in $F/M^2$.

Operatively, the method is extremely simple. The supersymmetry-breaking
gaugino masses, scalar masses, and coefficients of the $A$-type terms have
the expressions, derived in sect.~2,
\beq
{\tilde M}_g (\mu )=\left. -\frac{1}{2}\frac{\partial \ln S(X,\mu)}
{\partial \ln X}\right|_{X=M}~\frac{F}{M}~,
\label{gauginoc}
\eeq
\beq
{\tilde m}_Q^2(\mu )=\left. -\frac{\partial^2 \ln Z_Q(X,X^\dagger , \mu)}
{\partial \ln X ~\partial \ln X^\dagger}\right|_{X=M}
~\frac{FF^\dagger}{MM^\dagger}~,
\label{squakc}
\eeq
\beq
A_i(\mu )=\left. \frac{\partial \ln Z_{Q_i}(X,X^\dagger,\mu)}
{\partial \ln X}\right|_{X=M}~\frac{F}{M}~.
\label{aac}
\eeq
The expressions for the gauge and chiral wave-function renormalizations
$S$ and $Z_Q$ are obtained by integrating the well-known RG differential
equations. There is no need to evaluate any Feynman diagram. The method allows 
one to derive with no effort results which before required laborious
calculations. In particular, 
notice that the gauge-mediated
two-loop sfermion masses are obtained by integrating the
one-loop RG equation.

We have applied our method to derive several new results of phenomenological
interest. We have computed the sub-leading ${\cal O}(\alpha_t \alpha_s^2)$
corrections to the Higgs mass parameters in gauge-mediated models, suppressed
by a factor $1/\ln (M/{\tilde m}_t)$ with respect to the known leading
result. This correction modifies by at most 30 \% the value of the
Higgs mixing mass $\mu_H^2$ extracted from the electroweak symmetry-breaking
condition.

We have computed the supersymmetry-breaking terms induced by superpotential
couplings between matter and messenger chiral
superfields. These include $A$-type soft terms at one loop.
We have shown that an
interaction of this kind involving a Higgs singlet $N$ offers
a new possibility for a solution to the $\mu$ problem in gauge-mediated
theories. Indeed, the superpotential coupling between messengers and $N$,
generates a negative mass squared and trilinear couplings for the scalar
component of $N$, which can help to achieve the correct electroweak
breaking.

Finally, we have investigated the possibility that the messengers belong
to gauge supermultiplets instead of chiral supermultiplets, and calculated
the resulting soft terms. We have shown that
gauge messengers generate $A$-type trilinear terms at the messenger
scale, and give new contributions to scalar squared masses.
We have also shown that some recently-proposed models of 
dynamical supersymmetry breaking have contributions from gauge messengers
which drive negative slepton mass squared.
The presence of negative contributions to sfermion masses seems a generic
problem of models with gauge messengers. Since these models are otherwise
compelling, we believe it is interesting to search for realistic examples.

Since two-loop
$\beta$ functions and anomalous dimensions in supersymmetric theories
are know, our method seems well suited for next-to-leading order
calculations. However, we wish to warn the reader that to extend our method
beyond the leading order, one has to deal with some subtleties related
to the analytic continuation of the gauge wave-function~\cite{work}.

\bigskip

We wish to thank A.~Brignole,
S.~Dimopoulos, G.~Dvali, R.~Leigh, and M.~Luty for useful
discussions.

\def\ijmp#1#2#3{{\it Int. Jour. Mod. Phys. }{\bf #1~}(19#2)~#3}
\def\pl#1#2#3{{\it Phys. Lett. }{\bf B#1~}(19#2)~#3}
\def\zp#1#2#3{{\it Z. Phys. }{\bf C#1~}(19#2)~#3}
\def\prl#1#2#3{{\it Phys. Rev. Lett. }{\bf #1~}(19#2)~#3}
\def\rmp#1#2#3{{\it Rev. Mod. Phys. }{\bf #1~}(19#2)~#3}
\def\prep#1#2#3{{\it Phys. Rep. }{\bf #1~}(19#2)~#3}
\def\pr#1#2#3{{\it Phys. Rev. }{\bf D#1~}(19#2)~#3}
\def\np#1#2#3{{\it Nucl. Phys. }{\bf B#1~}(19#2)~#3}
\def\mpl#1#2#3{{\it Mod. Phys. Lett. }{\bf #1~}(19#2)~#3}
\def\arnps#1#2#3{{\it Annu. Rev. Nucl. Part. Sci. }{\bf #1~}(19#2)~#3}
\def\sjnp#1#2#3{{\it Sov. J. Nucl. Phys. }{\bf #1~}(19#2)~#3}
\def\jetp#1#2#3{{\it JETP Lett. }{\bf #1~}(19#2)~#3}
\def\app#1#2#3{{\it Acta Phys. Polon. }{\bf #1~}(19#2)~#3}
\def\rnc#1#2#3{{\it Riv. Nuovo Cim. }{\bf #1~}(19#2)~#3}
\def\ap#1#2#3{{\it Ann. Phys. }{\bf #1~}(19#2)~#3}
\def\ptp#1#2#3{{\it Prog. Theor. Phys. }{\bf #1~}(19#2)~#3}

\end{document}